\documentclass{aa}

\def\ha{H$\alpha$}
\def\hb{H$\beta$}

\def\lya{Ly$\alpha$}

\def\oii{[O~{\sc ii}]}
\def\oiii{[O~{\sc iii}]}

\def\hi{H~{\sc i}}

\usepackage{amsmath,bm}
\usepackage{caption}

\bibpunct{(}{)}{;}{a}{}{,}

\usepackage{graphicx}
\usepackage[varg]{txfonts}
\usepackage{hyperref}
\hypersetup{
  colorlinks   = true, 
  urlcolor     = blue, 
  linkcolor    = blue, 
  citecolor   = blue 
}
\usepackage[most]{tcolorbox}
\usepackage{xcolor}

\begin{document}

\title{BLUEPRINT: Blue-dominant Lyman-$\alpha$ emission as evidence of gas inflow in ultra-low-mass galaxies at $z=3$} \titlerunning{BLUEPRINT: gas inflow in ultra-low-mass galaxies at $z = 3$} \authorrunning{T. Mukherjee et al.}

\author{Tamal Mukherjee\inst{1}\fnmsep\inst{2}
          \and
          Zhihui Li \inst{3}
          \and
          Tayyaba Zafar\inst{1}\fnmsep\inst{2}
          \and
          Themiya Nanayakkara\inst{4}
          \and
          Davide Tornotti\inst{5}
          \and
          Luca Costantin\inst{6}
          \and
          Imam Uzma Aalia \inst{1}\fnmsep\inst{2}
          }
   \institute{School of Mathematical and Physical Sciences, Macquarie University, NSW 2109, Australia\\
   \email{tamal.mukherjee@hdr.mq.edu.au}
         \and
           Astrophysics and Space Technologies Research Centre,  Macquarie University, Sydney, NSW 2109, Australia
        \and
            Center for Astrophysical Sciences, Department of Physics \& Astronomy, Johns Hopkins University, Baltimore, MD 21218, USA
        \and 
           Centre for Astrophysics and Supercomputing, Swinburne University of Technology, P.O. Box 218, Hawthorn, 3122, VIC, Australia
        \and
            Dipartimento di Fisica ``G.~Occhialini'', Università degli Studi di Milano-Bicocca, Piazza della Scienza 3, 20126 Milano, Italy
        \and
            Centro de Astrobiología (CAB), CSIC-INTA, Ctra. de Ajalvir km 4, Torrejón de Ardoz 28850, Madrid, Spain
           }
   \date{Received XX, XXXX; accepted XX, XXXX}

\abstract
{We report the detection of a clumpy, blue-dominated Lyman-$\alpha$ (\lya) emission at $z = 3.066$ located in the heart of a cosmic web filament in the MUSE eXtremely Deep Field (MXDF).\ It is spatially associated with the formation of two compact star-forming regions revealed by deep JWST/NIRCam imaging. Gas accretion in these regions is indicated by the blue-dominated \lya\ profiles, spectral signatures that are rarely observed. Radiative transfer simulations of the \lya\ profile using a clumpy multiphase model suggest a radial inflow of gas clumps with a velocity of $\sim 100$ km/s. Embedded in this \lya\ structure, the associated main galaxy dominates the stellar mass budget, while the two compact ultra-low-mass systems ($\log (M_{\star}/M_{\odot})= 6.3 - 6.9$) have formed the bulk of their stellar mass in less than 7 Myr. These two components also have high specific star-formation rates and elevated ionisation parameters, consistent with recent bursty star formation. This system provides compelling observational evidence of how gas accretion, most likely from the cosmic web, can induce starbursts in ultra-low-mass galaxies.}
\maketitle


\section{Introduction}
\label{sec:int}

Galaxies build up their stellar mass over cosmic time by acquiring gas from multiple environmental reservoirs, which include large-scale structures such as the cosmic web \citep{Keres05, Dekel09}, recycled accretion of gas present in the circumgalactic medium (\citealt{Azar17}), and accretion from satellite and galaxy mergers \citep{Hafen19, Sparre22}. Despite the theoretical importance of gas accretion in driving galaxy evolution, especially around the peak of cosmic star formation (z $\sim$ 2–3), direct observational evidence remains sparse due to challenges such as the low covering fraction and surface brightness (SB) of cold, accreting neutral hydrogen (\hi ) gas filaments.

A powerful observational tracer of gas flows is Lyman-$\alpha$ (\lya) $\lambda 1215.67$ \AA\ emission. Due to its resonant nature, the \lya\ line is highly sensitive to the kinematics, geometry, and column density of neutral gas, providing key insights into gas exchange between galaxies and their surrounding environments. In most star-forming galaxies, \lya\ profiles are asymmetric and redshifted with respect to systemic velocity, a signature commonly interpreted as resonant scattering in outflowing gas.
In contrast, the scattering of \lya\ photons through infalling or accreting gas can produce blueshifted \lya\ peaks (e.g. \citealt{Dijkstra06}), leading to blue-dominated double-peaked \lya\ profiles (see \citealt{Mukherjee23},  \citealt{Bolda25}, and references therein). Such profiles are expected in theoretical models of gas accretion but are observationally rare. Only a small fraction ($\lesssim 20$ \%) of the overall population of double-peaked \lya\ emitters (LAEs) are found to exhibit a dominant blueshifted peak, both observationally \citep{Mukherjee25} and in cosmological zoom-in simulations \citep{Blaizot23}. A recent study of LAEs in the MUSE eXtremely Deep Field (MXDF), however, found a notable number of these systems \citep{Vitte25} but lacked systemic redshift measurements to confirm them as inflowing systems. Furthermore, blue-dominated profiles with a main peak followed by a smaller red bump could also arise from backscattering \citep{Mukherjee25}. Only a handful of blue-dominated LAEs with confirmed systemic redshifts are known to date \citep{Furtak22, Mc22, Mukherjee23, Bolda25}, confirming the presence of accreting gas and highlighting the difficulty of unambiguously identifying inflowing gas in emission.

\begin{figure*}[ht!]
\centering
    {\includegraphics[width=18cm,height=6cm]{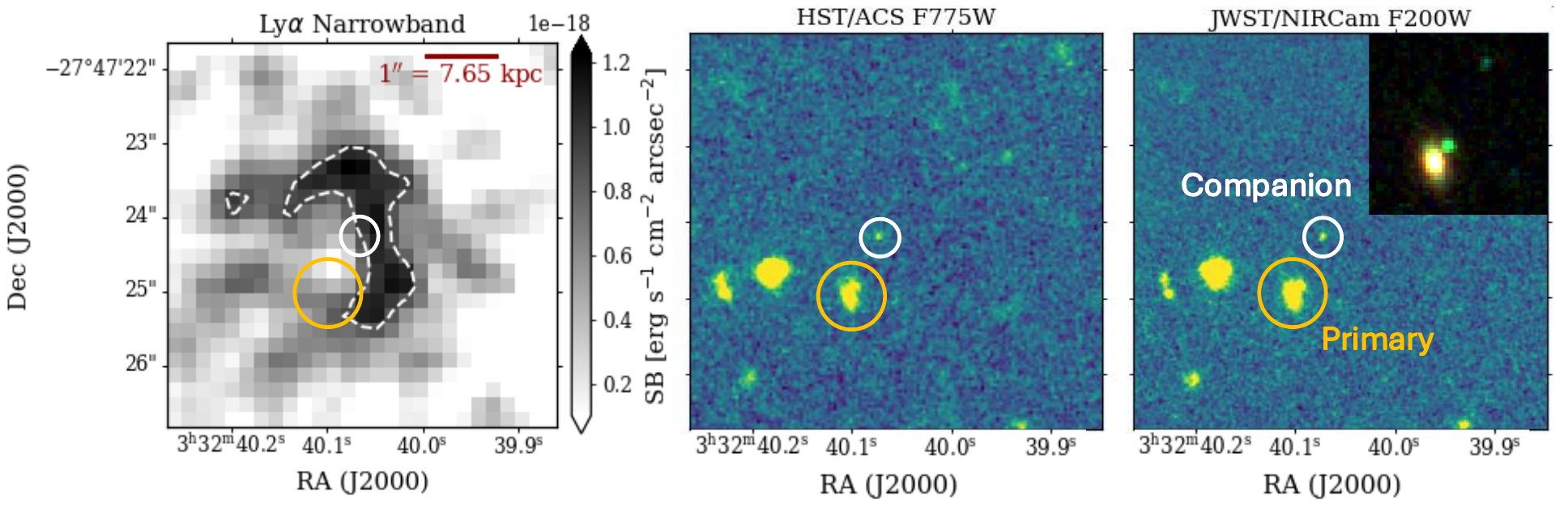}}
\caption{\textit{Left}: Continuum-subtracted \lya\ NB image obtained by collapsing the MUSE datacube over 4925–4960 \AA. The white dashed contour shows the 1.5 $\sigma$ significance level.  \textit{Middle}: HST/ACS F775W image. \textit{Right}: JWST/NIRCam F200W image. The yellow and white circles show the locations of the primary galaxy and the companion galaxy, respectively. The inset in the top-right corner of the right panel presents a zoomed-in RGB composite of the region containing the three galaxies, constructed using JWST/NIRCam F115W, F200W, and F356W filters. 
}
\label{fig:merged}
\end{figure*}

In this paper we report the detection of a clumpy blue-dominated \lya\ emission at $z = 3.066$ in the MXDF. It is embedded in a cosmic web filament \citep{Bacon21} in the \textit{Hubble} Ultra Deep Field and is spatially associated with two compact star-forming regions. By combining the Multi Unit Spectroscopic Explorer (MUSE) spectroscopy with deep JWST imaging, spectroscopy, and spectral energy distribution (SED) modelling, we found rare observational evidence of active gas accretion triggering star formation in ultra-low-mass galaxies. The structure of this paper is as follows. In Sect. \ref{sec:data} we provide details of the observations and data. Sect. \ref{sec:methods} presents the methods carried out to analyse the data. In Sect. \ref{sec:discussion} we present our results, followed by a discussion. Finally, we summarise our results in Sect. \ref{sec:summary}. Throughout this paper, we assume a standard flat $\Lambda$ cold dark matter cosmology with parameters $H_0$= 70 $\mathrm{km} \, \mathrm{s}^{-1} \mathrm{Mpc}^{-1}$, $\Omega_{\mathrm{m}}$ = 0.3, and  $\Omega_{\Lambda}$ = 0.7.

\section{Data}
\label{sec:data}

\subsection{MUSE integral field spectroscopy}

The primary galaxy and the \lya\ emission are detected in the MUSE integral field spectroscopic data from the MXDF, a 140 h deep MUSE observation located in the \textit{Hubble} Ultra-Deep Field \citep{Bacon21, Bacon23}. The MXDF observations were obtained as part of the MUSE Guaranteed Time Observations programme between August 2018 and January 2019, and were carried out using the MUSE ground-layer adaptive optics mode. These are the deepest MUSE data ever obtained, allowing the detection of faint LAEs down to luminosities of $\sim 10^{40}$ $\mathrm{erg}\, \mathrm{s}^{-1}$ at $z \sim 3$. Details of MXDF data reduction and source catalogue construction are provided in \citet{Bacon23}.
In the MXDF, the full width at half maximum of the Moffat point-spread function \citep[PSF;][]{Moffat69} is approximately 0.6$''$ at 4700~\AA\ and 0.4$''$ at 9350~\AA. The line spread function is spatially constant across the field but increases towards the outer regions. The mean MXDF line spread function  over the full wavelength range is 2.6~\AA. 

The corresponding blue-dominated double-peaked \lya\ spectrum \citep[MUSE ID: 8360 in the Data Release 2 catalogue,][]{Bacon23} is presented in the \lya\ spectral classification study by \cite{Vitte25}. No other emission line is seen in MUSE.
The trough between the two \lya\ peaks yields a redshift of $z = 3.066$. The continuum-subtracted \lya\ narrowband (NB) image reveals an extended clumpy structure of \lya\ (see the left panel of Fig. \ref{fig:merged}). Using the publicly available MUSE data products\footnote{\href{https://amused.univ-lyon1.fr/}{https://amused.univ-lyon1.fr/}}, we identify a spatial offset of $0.66''$ between the primary galaxy (MUSE ID: 8497) and the location of maximum \lya\ SB. Furthermore, this \lya\ system is embedded in a cosmic web filament \citep[4 cMpc extended,][]{Bacon21}.

\subsection{HST Imaging}

We utilised data from the \textit{Hubble} Space Telescope (HST) imaging. The data include imaging obtained with the Advanced Camera for Surveys (ACS) in the four optical filters F435W, F606W, F775W, and F850LP, providing deep rest-frame ultraviolet–to–optical coverage at the redshift of our source. The HST data reveal a faint clump or companion galaxy located at a projected spatial offset of $0.85''$ ($\sim$ 6.5 kpc at that redshift) from the primary galaxy (see the middle panel of Fig. \ref{fig:merged}), while the companion itself is offset by $0.2''$ from the \lya\ SB peak. 
Both the primary galaxy (RAF\_ID = 5199) and C (RAF\_ID = 5177) are included in the photometric catalogue of \cite{Rafelski15}. The uniform depth and high spatial resolution of the ACS imaging allow robust photometric measurements and morphological characterisation of both components.

\subsection{JWST NIRCam and MIRI Imaging}

We further combined deep JWST/NIRCam observations of the Great Observatories Origins Deep Survey–South field (GOODS-S) from the JWST Advanced Deep Extragalactic Survey \citep[JADES;][]{Bunker24, Rieke23, Eisenstein23} and the FRESCO survey \citep{Oesch23}, providing extensive broad and medium band coverage.
 Within this region there are imaging data from 14 NIRCam filters: F090W,
F115W, F150W, F182M, F200W, F210M, F277W, F335M, F356W, F410M, F430M, F444W, F460M, and F480M. Both sources are detected in the NIRCam filters. Further, 
we identified another resolved component (primary-1 hereafter) associated with the primary galaxy in some of the medium- and broad-band filters. The red-green-blue (RGB) composite (see the right panel of Fig~\ref{fig:merged}) reveals clear colour differences among the three components. The primary galaxy appears redder, whereas primary–1 and the companion appear significantly greener, showing intense nebular line (\oiii) emission relative to continuum. We discuss the resolved photometric extraction in Sect. \ref{sec:phot_resolved}.

\begin{figure*}[ht!]
\centering
    {\includegraphics[width=18cm,height=5.5cm]{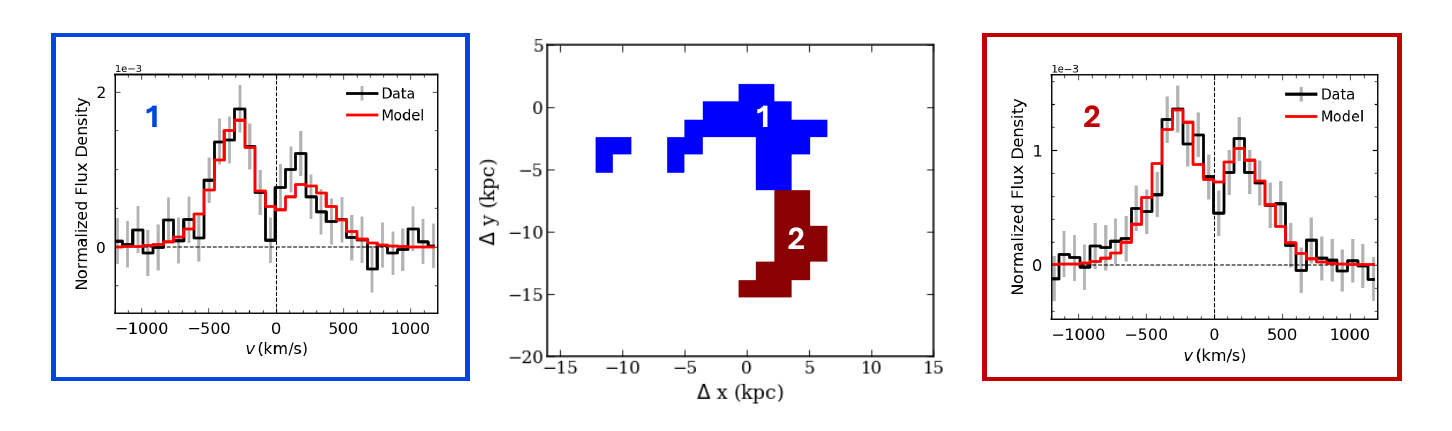}}
\caption{Radiative transfer modelling of spatially resolved \lya\ profiles (\textit{left} and \textit{right} panels) extracted from Voronoi-binned regions (\textit{middle} panel), along the direction 1 $\rightarrow$ 2. The observed line profiles are shown in black (with uncertainties shown in grey), and the best-fit models
are shown in red. The vertical dashed line represents the line centre velocity corresponding to a systemic redshift of $z = 3.066$, obtained from JWST/NIRSpec spectra. The middle panel shows two Voronoi-binned regions displayed on the \lya\ NB image within a 1.5 $\sigma$ significance, centred on the pixel with maximum \lya\ SB. 
}
\label{fig:modelling}
\end{figure*}

Additionally, we included JWST/ Mid-Infrared Instrument (MIRI) imaging data obtained as part of the MIRI \citep{Rieke15, Bouchet15, Wright23, Dicken24} Deep Imaging Survey (MIDIS; PID: 1283) of the \textit{Hubble} Ultra Deep Field \citep{Ostlin25}. This survey provides some of the deepest mid-infrared observations, enabling sensitive probes of the rest-frame near-infrared emission from distant galaxies \citep{Costantin25}. The primary galaxy is detected in two MIRI filters (F560W and F770W), while both primary-1 and the companion remain undetected in the MIRI images.

\subsection{JWST NIRSpec spectroscopy}

The JWST/NIRSpec spectrum of the primary galaxy was obtained as part of the NIRCam–NIRSpec Galaxy Assembly survey of GOODS-S (proposal ID 1286; PI: Luetzgendorf). This programme includes both PRISM spectroscopy and medium- and high-resolution grating--filter combinations, spanning the full wavelength range of 1–5 $\mu$m. In this work, we used the spectrum obtained with the medium resolution grating--filter combination, which covers all of the strongest nebular emission lines of interest at medium spectral resolution ($R \sim 1000$), making it well suited for systemic redshift determination. Fitting Gaussian profiles to the \oiii\ and \ha\ emission lines, we measured a systemic redshift of $z_{\mathrm{primary}} = 3.0662$. This redshift is consistent with the spectroscopic redshift reported in the JADES DR4 catalogue \citep{Curtis25, Scholtz25} for the primary galaxy (NIRSpec ID: 118073). The slit was positioned such that it covers both the primary galaxy and primary-1 (see Fig.~\ref{fig:NIRSpec_spectrum}, top panel). However, we did not detect any additional emission lines associated with primary-1 in either the PRISM or grating spectra. We therefore infer that primary-1 is at the same redshift as the primary galaxy.
The spectrum of the companion was obtained recently as part of the proposal ID 8060 (PI: Egami), with source ID 118266. Only a PRISM spectrum is available for this source. From this, we estimated a systemic redshift of $z_{\mathrm{companion}} = 3.066$. This confirms that the companion lies at the same redshift as the primary galaxy, indicating that it is part of the same system. The NIRSpec spectra are presented in Fig.~\ref{fig:NIRSpec_spectrum}. These data are available at Mikulski Archive for Space Telescopes (MAST) and can be accessed using the DOI: \href{https://doi.org/10.17909/qy9s-8f72}{https://doi.org/10.17909/qy9s-8f72}.

\section{Methods and analysis}\label{sec:methods}

Based on MUSE, HST, and JWST observations, this system appears to be a blue-peak dominated \lya\ emission surrounding three galaxies (primary, primary-1, and companion) and indicating gas accretion in multiple spatial locations. In this section we present methods for studying this intriguing system. 

\subsection{Spatially resolved \lya\ }

We extracted a \lya\ spectrum by summing all spaxels with S/N $> 1.5$ in the MUSE datacube and fitted a double-asymmetric Gaussian function similar to \cite{Mukherjee25}. We obtained a total \lya\ flux of $(3.97 \pm 0.31) \times 10^{-18}$ $\mathrm{erg}\, \mathrm{cm}^{-2}\, \mathrm{s}^{-1}$. Because of the faintness of the emission, we further implemented a Voronoi binning technique  to
subdivide the halo into regions with
reasonable S/N. We performed Voronoi binning on
spaxels with S/N > 1.5 in the halo, as shown
in the \lya\ NB image. This procedure yielded two Voronoi-binned regions (Fig.~\ref{fig:modelling}) with \lya\ S/N in the range 7.5 - 10. We then extracted the \lya\ spectra from each Voronoi-binned
region by summing the continuum-subtracted datacube
across all spaxels. The spectra from both bins exhibit a blue-dominant double-
peaked \lya\ profile, indicating gas inflows.  We measured a blue-to-total flux ratio of $\mathrm{B/T} \sim 0.65$ in both bins.

\subsection{Radiative transfer modelling using a multiphase clumpy model}

We modelled the spatially resolved \lya\ profiles with a multiphase clumpy radiative transfer model as outlined in \cite{Li21, Li22} and \cite{Bolda25}. In this model, cool ($\sim 10^{4}$ K), spherical \hi\ clumps are assumed to be embedded in and move within a hot ($\sim 10^{6}$ K), ionised inter-clump medium (ICM). The model is characterised by five free parameters: the clump volume filling factor ($F_{\mathrm{v}}$), which is related to cloud covering factor $f_{cl}$; the residual \hi\ number density in ICM ($\mathrm{n}_{\mathrm{HI, ICM}}$), which determines the depth of the absorption trough; the velocity
dispersion of the clumps ($\sigma_{\mathrm{cl}}$); a constant radial velocity $v_{\mathrm{cl}}$; and the \hi\ column density of the clumps ($N_{\mathrm{HI, cl}}$). The best-fit model for the spectra from the two Voronoi binned regions are shown in Fig. \ref{fig:modelling}, and the corresponding combined best-fit parameters are listed in Table \ref{tab:modelling}. The best-fit model suggests a clump radial inflow velocity of $v_{\mathrm{cl}} \sim 98 $ km/s. 

\begin{table}[ht!]
    \centering
    \caption{Best-fit parameters of \lya\ radiative transfer modelling.}
    \begin{tabular}{lc}
       Parameters  & Best-fit value  \\ \hline \hline \\ [-1pt] 
       $v_{\mathrm{cl}}$ \, ($\mathrm{km} \, \mathrm{s}^{-1}$) & $-97.82^{+ 49.40}_{-48.42}$  \\ [6pt]       
       $F_{\mathrm{V}}$ & $0.112^{+ 0.03}_{-0.04}$ \\ [6pt]
       log ($n_{\mathrm{HI, ICM}} / \mathrm{cm}^{-3}$) & $- 7.53^{+ 0.12}_{-0.23}$ \\ [6pt]
       log ($N_{\mathrm{HI, cl}} / \mathrm{cm}^{-2}$) & $17.84^{+ 0.35}_{-0.31}$ \\ [6pt]
       $\sigma_{\mathrm{cl}}$ \, ($\mathrm{km} \, \mathrm{s}^{-1}$) & $226.71^{+ 22.03}_{-39.44}$ \\ [3pt]
       \hline
    \end{tabular}
    \label{tab:modelling}
\end{table}

\subsection{Photometric extraction and $\beta_{UV}$ estimation}\label{sec:phot_resolved}

For companion, we extracted photometry by constructing segmentation maps using the JADES NIRCam/F200W image with \texttt{Photutils} \citep{Bradley25}. The F200W image was adopted as the detection image because it provides the highest signal-to-noise among the available NIRCam observations, enabling robust source segmentation. Local background properties were measured, and the median background was subtracted prior to source detection. We employed the following segmentation parameters for source detection: threshold = 2.5 $\sigma$ per pixel above the background noise, npixels=10. Kron aperture for segment C is set with parameter of 2.0 (KRON). We applied same Kron aperture consistently across all filters, prioritising the deeper JADES
images over the shallower FRESCO images. 

\begin{figure}[ht!]
    \centering
    \includegraphics[width=0.9999\linewidth]{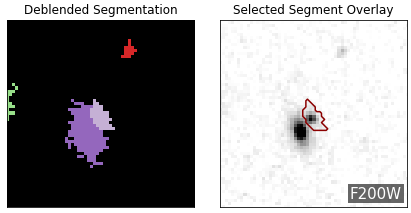}\\
    \includegraphics[width=0.9999\linewidth]{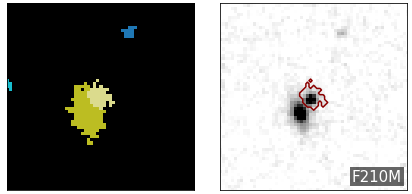}
    \caption{De-blended segmentation map of the NIRCam F200W and F210M filter showing the segmented region for primary-1 along with the primary galaxy.}
    \label{fig:deblending}
\end{figure}

To extract resolved photometry for clump primary-1, which overlaps with the primary galaxy, we employed a de-blended segmentation approach based on a multi-threshold watershed algorithm. De-blended segmentation was performed on the NIRCam/F200W and F210M filters with the  following parameters: npixels = 6, nlevels = 10, and contrast = 0.001. Figure~\ref{fig:deblending} shows the de-blended segmentation maps highlighting the identified segment corresponding to primary-1 in two NIRCam filters. We adopted the deeper F210M image to define the segmentation and obtain resolved photometry by summing the pixel values within the target segment. To ensure consistent photometry across all bands, all other photometric bands are PSF-matched to the F210M PSF and re-projected onto the F210M pixel grid. Fluxes are then measured using identical segmentation masks across all filters for the primary and primary-1. 

We estimated the UV continuum slope, $\beta_{UV}$ (see Table \ref{tab:SED}), by fitting $f_{\lambda} \propto \lambda^{\beta}$ to the photometry sampling rest-frame $1250-2600$ \AA. We obtained a bluer UV slope, $\beta_{UV}$ $= -1.74 \pm 0.19$, and $-1.56 \pm 0.67$ for primary-1 and companion, respectively, indicating younger stellar populations.

\subsection{Spectral energy distribution modelling using Bagpipes}\label{sec:SED}

We performed SED fitting on the extracted photometric data using the code Bayesian Analysis of Galaxies for Physical Inference and Parameter EStimation \citep[BAGPIPES;][]{Carnall18}. The model uses a \cite{Kroupa2001} initial mass function and implements the \cite{BC03} stellar population models. A double power-law star formation history (SFH) is adopted, which provides sufficient flexibility to describe both rising and declining SFHs while avoiding the strong assumptions inherent to simpler parametric forms such as exponentially declining or constant SFHs. The peak time parameter ($\tau$) was allowed to vary between the Big Bang at 0 Gyr and 15 Gyr later. The total stellar mass formed was sampled over the interval 
$\log(M/M_{\odot}) \in [1, 15]$, while the stellar metallicity was allowed to vary within 0 to
$2.5\, Z_{\odot}$. The dust attenuation model of \cite{Calzetti2000} was used, with the visual extinction allowed to vary within the range $0 < A_V < 2$. Nebular emission was included self-consistently, allowing the ionisation parameter to vary over 
logU= -4 to -1. We additionally allowed the escape fraction of ionising photons, 
$f_{\mathrm{esc}}$, to vary between 0 and 1. This conservative choice accounts for uncertainties in the coupling between the ionising stellar population and the surrounding gas, particularly in low-mass systems where ionising photon leakage may be significant. The physical properties inferred from the SED fitting are presented in Table \ref{tab:SED}, and the corresponding best-fit SEDs are shown in Fig.~\ref{fig:app2}. For primary-1, for which no spectroscopic redshift is available, we obtained a photometric redshift of $z = 3.078^{+0.017}_{-0.014}$, consistent with the systemic redshift of the primary galaxy and supporting our assumption that it is part of the same system.

\begin{table*}[ht!]
\centering
\caption{Physical properties of the galaxies.}
\begin{tabular}{lccccccccccc}
Galaxy & $\beta_{UV}$ & log $(M_{*}/M_{\odot})$ & $\langle t\rangle_{\mathrm{MW}}$ (Myr) & $\mathrm{SFR}_{\mathrm{SED}}$ ($M_{\odot}$ / yr) & $\mathrm{SFR}_{H\alpha}$ ($M_{\odot}$ / yr) & 12 + log(O/H) & log $U$ \\ \hline \hline \\ [-1pt] 
Primary & $-1.95 \pm 0.21$ & $8.39^{+0.31}_{-0.12}$ &  80 & $1.78^{+0.34}_{-0.29}$ & 0.79 $\pm$ 0.06 & $8.17 \pm 0.51$ & $-3.059^{+0.178}_{-0.115}$ \\ \\ [1pt]
Primary-1 & $ -1.74 \pm 0.19$ & $6.91^{+0.22}_{-0.21}$  &  5 & $0.086^{+0.007}_{-0.006}$ & ... & ... & $-2.316^{+0.299}_{-0.315}$\\ \\ [1pt]
Companion  & $-1.56 \pm 0.67$ & $6.28^{+0.23}_{-0.18}$  &  7 & $0.023^{+0.003}_{-0.003}$ & $0.18 \pm 0.03 $ & $7.16 \pm 0.38$ & $-1.584^{+0.358}_{-0.340}$\\ [2pt]
\hline
\end{tabular}
\label{tab:SED}
\end{table*}

\subsection{Emission line diagnostics}

For the primary galaxy, strong emission lines such as \hb $\lambda 4861$, the \oiii $\lambda 4959, 5007$ doublet, and \ha $\lambda 6563$ are robustly detected at S/N $> 5$ in G235M F170LP filter configuration. In addition, \oii $\lambda 3728$ is detected in G140M F070LP filter. We adopted emission-line flux measurements from the publicly available JADES NIRSpec DR4 emission-line catalogue \citep{Curtis25, Scholtz25}. Due to the absence of temperature-dependent auroral emission lines such as \oiii $\lambda 4363$, we are unable to directly constrain the gas-phase metallicity. We therefore instead employed strong-line metallicity diagnostics. Following \cite{Sanders24}, we used empirical calibrations based on the O3 $\equiv$ \oiii\ 5007/\hb\ line ratios, in which the gas-phase metallicity is parameterised as a polynomial function of these strong-line ratios. To account for the intrinsic double-valued nature of the O3 diagnostic, we additionally verified the solution using the single-valued O32 calibration. Both diagnostics consistently yield a gas-phase metallicity of 12 + log(O/H) $= 8.17 \pm 0.51$, where the uncertainty includes both emission-line flux uncertainties and intrinsic calibration scatter.

Additionally, we utilised the prism spectrum of the companion galaxy to obtain line fluxes by fitting Gaussian profiles. Due to the faintness of the source, only H$\alpha$ and [O\,\textsc{iii}]~$\lambda5007$ emission lines are robustly detected, while H$\beta$ is not directly resolved. We therefore inferred the H$\beta$ flux assuming a standard Case~B recombination Balmer decrement (\ha/\hb = 2.86). Using the inferred \oiii/\hb\ ratio, we computed the O3 index and applied the same metallicity calibration as adopted for the primary galaxy. The O3 diagnostic yields two degenerate metallicity solutions. Given the ultra-low stellar mass derived from SED fitting, the high-metallicity solution is unlikely in the context of the galaxy mass--metallicity relation. We therefore adopted the low-metallicity branch, corresponding to $12+\log({\rm O/H}) = 7.16 \pm 0.38$.

From \ha\ emission-line fluxes, we derived star-formation rates (SFRs) using the calibration of \cite{Kennicutt98}, converted to a \cite{Kroupa2001} initial mass function for consistency with the SED modelling. For the primary galaxy, the observed \ha\ flux is first corrected for dust attenuation using the Balmer decrement, while no dust correction was applied for the companion due to the lack of H$\beta$ measurements. The \ha-based SFR values  ($\mathrm{SFR}_{\mathrm{H}\alpha}$) are listed in Table \ref{tab:SED}.

\section{Results and discussions}
\label{sec:discussion}

This paper presents the discovery of clumpy blue-dominant \lya\ emission surrounding galaxies at $z = 3.066$. 

\subsection{\lya\ as evidence of clump radial inflow}

Radiative transfer modelling of blue-dominated \lya\ profile using the clumpy multiphase model predicts a strong gas inflow with a velocity of $\sim 98$ km/s. This is consistent with theoretical predictions, in which a dominant blueshifted \lya\ peak naturally arises when photons preferentially escape through an infalling neutral medium \citep{Dijkstra06, Verhamme06}. Importantly, this system is located in the heart of a large-scale cosmic web filament, where continuous baryonic inflow from intergalactic filaments is theoretically expected \citep{Keres05, Dekel09, FG11}. The derived volume filling factor $F_V = 0.11$ implies a covering factor $f_{cl} \simeq 8.4$ for clumps of radius = 100 pc within a simulation sphere of 10 kpc radius. This corresponds to an \hi\ covering fractions effectively close to unity, indicating that nearly all sightlines intersect neutral clumps. This result is consistent with previous observational studies at $z > 3$ suggesting that the \hi\ covering fraction around galaxies with low \lya\ SB is close to unity \citep[e.g.][]{Wisotzki18}. Further, the extremely low residual inter-clump neutral density, log ($n_{\mathrm{HI, ICM}} / \mathrm{cm}^{-3}$) $\simeq -7.5$, indicates that most of the neutral opacity resides in discrete clumps rather than in a volume-filling diffuse phase. Therefore, these results support a picture in which the neutral circumgalactic medium is highly fragmented into compact clumps embedded within a mostly ionised medium.

\subsection{Stellar populations and star-formation activities}

From the SED best-fitting parameters, we find that the primary galaxy dominates the stellar mass budget ($\log M_{\star}/M_{\odot}=8.39$), while primary-1 and the companion are $\sim 3.3\%$ and $\sim 0.9\%$ of the stellar mass of the primary galaxy, respectively. 
The primary galaxy has $\mathrm{SFR}_{\mathrm{SED}}=1.78\,M_{\odot}\,\mathrm{yr}^{-1}$, implying a specific star-formation rate $(\mathrm{sSFR}) \equiv \mathrm{SFR}_{\mathrm{SED}}/M_{\star}\simeq 10^{-8.14}\,\mathrm{yr}^{-1}$. In contrast, primary-1 and the companion occupy a more extreme regime. Primary-1 has $\log (M_{\star}/M_{\odot})=6.91$, $\mathrm{SFR}_{\mathrm{SED}}=0.086\,M_{\odot}\,\mathrm{yr}^{-1}$, and mass-weighted age $\langle t\rangle_{\mathrm{MW}}= 5$~Myr.
The companion is similarly young and bursty, with $\log (M_{\star}/M_{\odot})=6.28$, $\mathrm{SFR}_{\mathrm{SED}}=0.023\,M_{\odot}\,\mathrm{yr}^{-1}$, $\langle t\rangle_{\mathrm{mw}}= 7$~Myr. SFRs derived from different diagnostics probe different timescales. While SED-based SFR traces star formation averaged over longer periods, H$\alpha$ traces much more recent star formation on timescales of $\sim5$–$10$ Myr. 

For companion, quantitatively we find $\mathrm{SFR}_{\mathrm{H}\alpha}/\mathrm{SFR}_{\mathrm{SED}} \sim 8$, indicating that the recent ($\lesssim10$ Myr) star formation activity significantly exceeds the longer-timescale average.
Furthermore, both primary-1 and companion exhibit very high $\mathrm{sSFR}\simeq 10^{-7.98}\,\mathrm{yr}^{-1}$, firmly placing them in the starburst regime \citep{Rinaldi25}. These sSFRs imply rapid growth (mass-doubling in $\sim 0.1$~Gyr) and strongly support the interpretation that primary-1 and the companion are experiencing very recent and likely ongoing star-formation episodes. 

The inferred ionisation parameters also show a strong progression across the system. The primary has a relatively modest $\log U=-3.059$, whereas primary-1 has $\log U=-2.316$ and the companion reaches $\log U=-1.584$. Higher $U$ corresponds to a larger ionising photon density relative to the gas density, and is commonly associated with compact star-forming regions hosting very young massive stars. The combination of extremely young ages and high $\log U$ in primary-1 and the companion therefore suggests harder and more intense local ionising radiation fields than in the primary's integrated light.
Importantly, these two galaxies are co-spatial with regions that exhibit strong blue-peaked \lya\ emission. The coincidence of (i) blue-dominant \lya\ profiles and inferred inflow kinematics from modelling, with (ii) very young stellar populations, high sSFR, and elevated ionisation parameters, strongly supports a scenario in which active gas accretion is fuelling and triggering the current star-formation episodes in the compact ultra-low-mass substructures. The low gas phase metallicity of the companion further indicates a chemically young and metal-poor interstellar medium consistent with a rapidly assembling low-mass system.

\subsection{NFW density profile modelling}

To further understand the physical connection between the three galaxies, we examined their dark matter environment using Navarro–Frenk–White (NFW)  density profile modelling \citep{Navarro97}. In this framework, the primary galaxy and the compact primary–1 component are treated as a single dominant system defining the central halo, while the companion is modelled as an independent sub-halo placed at a projected spatial offset of 6.5 kpc. This approach allows us to assess whether the companion is gravitationally embedded within the primary’s halo and whether it can remain dynamically distinct.

\begin{figure}[ht!]
    \centering
    \includegraphics[width=0.9999\linewidth]{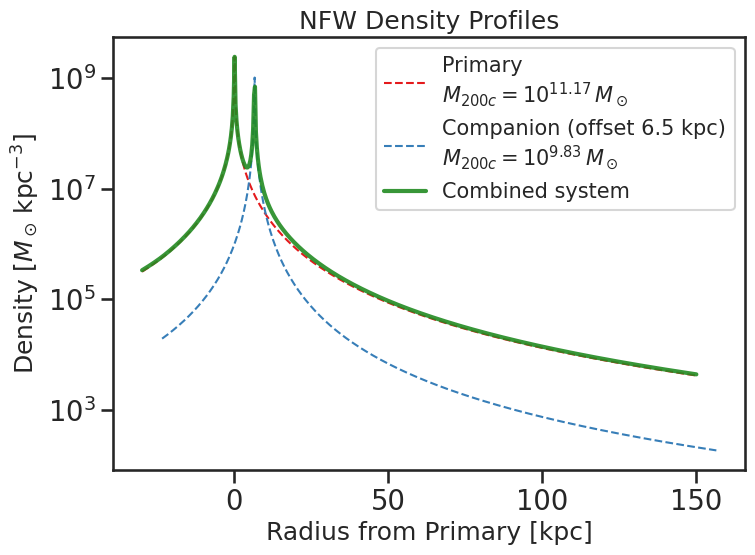}
    \caption{NFW dark matter density profiles for the system at $z= 3.066$. The dashed red curve shows the NFW profile of the primary galaxy halo (including the compact primary–1 component), the dashed blue curve shows the companion modelled as a sub-halo placed at a projected offset of 6.5 kpc, and the solid green curve shows the combined density profile.}
    \label{fig:NFW}
\end{figure}

To construct the NFW profiles, we first estimated the dark matter halo masses associated with the galaxies by converting their stellar masses into halo masses using the stellar–to–halo mass relation of \cite{Moster13}. We note that the stellar masses of the compact components (primary-1 and companion) lie near or below the low-mass regime over which empirical stellar-to-halo mass relations are well constrained. Consequently, the inferred halo masses for these systems should be interpreted with caution. The halo mass $M_{200c}$ is defined as the mass enclosed within a radius $R_{200c}$ where the mean density is 200 times the critical density of the Universe at the systemic redshift. For each halo, the concentration parameter is computed using the redshift-dependent concentration–mass relation implemented in the \texttt{Colossus} package \citep{Diemer19}. The corresponding scale radius $r_s = R_{200c}/c_{200c}$ and characteristic density $\rho_s$ were then derived from the standard NFW parametrisation. The combined profile is computed by shifting the companion NFW by 6.5 kpc and summing it with the primary NFW at each radius on a common radial grid. The resulting NFW density profiles (Fig. \ref{fig:NFW}) show that the companion is clearly located well inside the virial region of the primary halo. The combined density profile is dominated by the primary at all radii, confirming that the primary galaxy defines the overall gravitational potential of the system. However, the companion’s own NFW profile remains well defined and locally distinct, indicating that it survives as a coherent sub-halo rather than being fully disrupted or dissolved into the central halo. This configuration is broadly consistent with an early-stage satellite–host interaction within a common dark matter halo.

\subsection{Accretion from the cosmic web filament}

As mentioned before, the system lies within a large-scale cosmic web filament, where diffuse \hi\ gas is expected to flow along filamentary structures and accrete onto galaxies. The prominent blue peaks in the \lya\ profiles and the inflow velocities inferred from modelling are very consistent with this picture. In general, low-metallicity gas from the cosmic web supplies fresh fuel for star formation and galaxy growth at high redshift \citep{Keres05, Dekel09}. While shock-heated hot-mode accretion of metal-enriched gas may dominate in more massive systems with $\log(M_\star)\gtrsim10.5$ at cosmic noon \citep{Dekel09, Bolda25}, our system hosts low-mass galaxies where relatively cool, metal poor gas can efficiently accrete along filamentary streams without being shock heated, providing fresh fuel for star formation. 
Within this framework, the primary galaxy likely resides near the centre of the shared dark matter halo and defines the overall gravitational potential of the system. Its older mass-weighted age, higher metallicity, and lower ionisation parameter suggest a more evolved stage of star formation. In contrast, the compact components primary–1 and the companion may trace localised star-forming regions triggered by cold gas streams feeding the halo. A similar scenario has recently been suggested by \cite{Tornotti25}, who identified LAEs embedded within a cosmic web filament traced by extended \lya\ emission in the MUSE Ultra Deep Field \citep{Lusso19}. Some of their sources exhibit blue-dominated double-peaked \lya\ profiles, interpreted as possible signatures of gas accretion from the surrounding intergalactic medium, consistent with the picture suggested for our system.

\subsection{Recycled accretion?}
An alternative possibility is that some fraction of the accreting gas could be recycled material \citep{Oppenheimer10, Azar17}. Outflows driven by earlier star formation in the primary galaxy may cool, lose momentum, and subsequently fall back towards the system, where they can fuel new star formation. In this scenario, the recycled gas is expected to preferentially accrete and settle into local gravitational potential minima, such as those associated with compact dark matter sub-halos. This potentially explains their spatial association with the primary galaxy and their elevated star-formation activity. However, with the currently available data we cannot uniquely distinguish between pristine gas accretion from the cosmic web filament and recycled accretion, and both processes may contribute to the gas supply in this system.

\section{Summary}
\label{sec:summary}

Our system provides a unique window into the dynamic processes governing early galaxy formation and evolution. By leveraging the unprecedented capabilities of JWST, we captured a critical phase of galactic assembly, revealing how interactions and gas accretion can trigger star formation in ultra-low-mass galaxies and drive complex gas dynamics. The detection of strong gas inflows within the \lya\ filament, together with the emergence of compact star-forming regions, highlights the intricate mechanisms through which galaxies grow at high redshift.

Overall, this system offers a compelling observational example of gas accretion operating across multiple spatial scales within a single dark matter halo, simultaneously sustaining galaxy-wide growth while triggering intense, localised star formation in low-mass components. In this context, this system establishes blue-dominant \lya\ emission as a direct probe of cosmic-web-driven baryon inflow and rapid stellar mass growth in faint galaxies. Future high-resolution studies, particularly those combining multi-wavelength observations with advanced radiative transfer modelling, will be essential for further disentangling the interplay between galaxy interactions, gas accretion, and star formation in the early Universe.

\begin{acknowledgement}
We thank the anonymous referee for their constructive feedback. TM acknowledges support through the International Macquarie Research Excellence Scholarship Program (iMQRES).
TN thanks support through ARC Discovery Project Grant DP230103161. The project that gave rise to these results received the support of a fellowship from the “la Caixa” Foundation (ID 100010434). The fellowship code is LCF/BQ/PR24/12050015.  This work is based on
observations taken by VLT, which is operated by European Southern Observatory. This work utilises observations made with the NASA/ESA/CSA \textit{James Webb} Space Telescope. The data were obtained from the Mikulski Archive for Space Telescopes at the Space Telescope Science Institute, which is operated by the Association of Universities for Research in Astronomy, Inc., under NASA contract NAS 5-03127 for JWST.

\end{acknowledgement}

\bibliographystyle{aa} 
\bibliography{aanda} 

\appendix

\section{JWST/NIRSpec spectra}

\begin{center}
\includegraphics[width=\columnwidth]{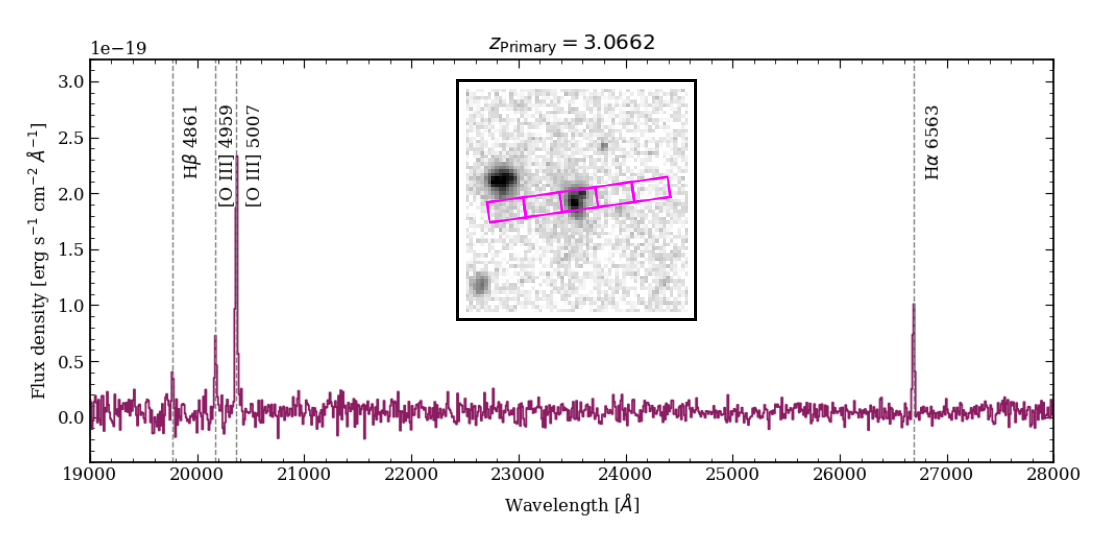}\\[0.3cm]
\includegraphics[width=\columnwidth]{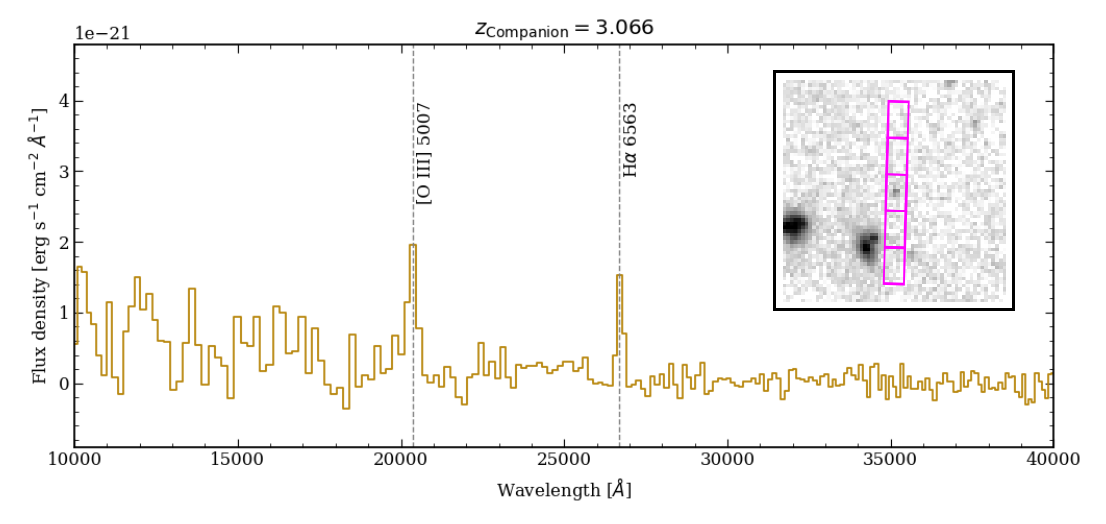}

\captionof{figure}{\textit{Top panel}: JWST/NIRSpec medium-resolution G235M F170LP spectrum obtained at the location of the primary galaxy (NIRSpec ID = 118073). \textit{Bottom panel}: Low-resolution PRISM spectrum of the companion galaxy.
Insets show the corresponding slit positions in each panel.}
\label{fig:NIRSpec_spectrum}
\end{center}


\section{Best-fit SEDs}

\begin{center}
\includegraphics[width=\columnwidth]{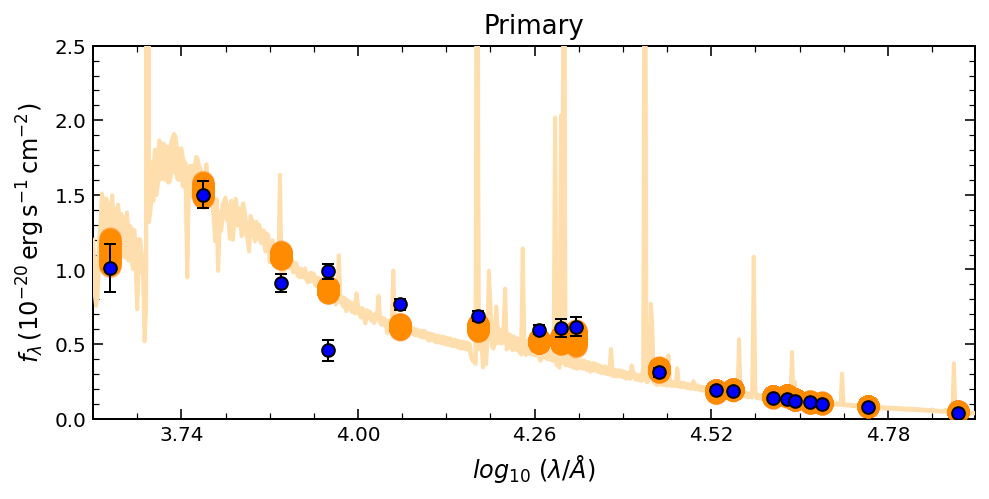}\\[0.3cm]
\includegraphics[width=\columnwidth]{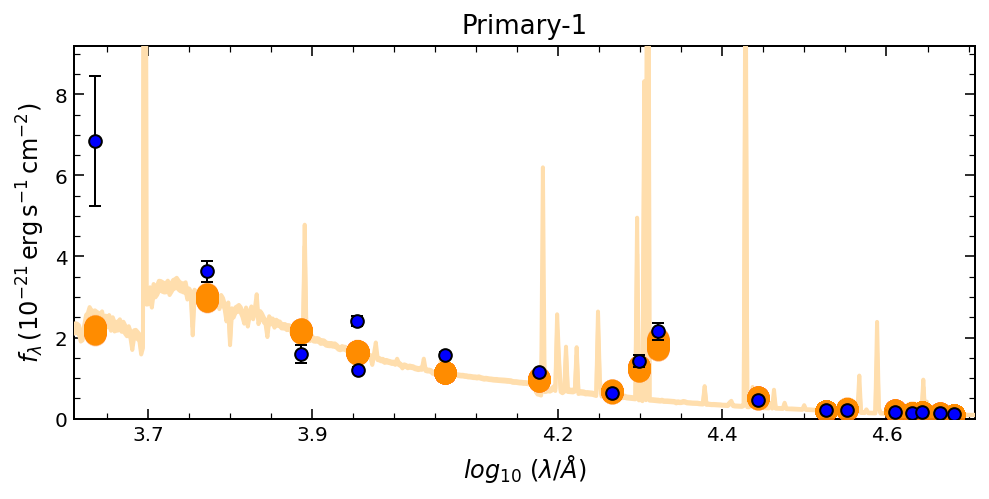}\\[0.3cm]
\includegraphics[width=\columnwidth]{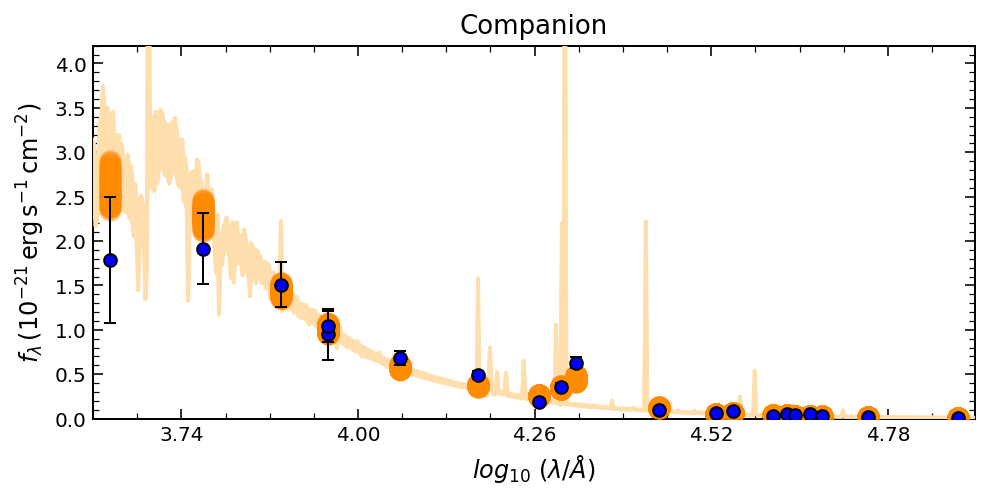}

\captionof{figure}{
Best-fit SED models for the primary, primary-1, and companion galaxies.
}
\label{fig:app2}
\end{center}

\end{document}